\documentclass{article}
\usepackage{spconf,amsmath,epsfig,amssymb,xcolor, float,url}

\newcommand{\speckle}{{u}}
\newcommand{\reflectivity}{{v}}
\newcommand{\intensity}{{w}}

\newcommand{\ratio}{{\tau}}
\newcommand{\superimage}{{s}}

\title{FAST STRATEGIES FOR MULTI-TEMPORAL SPECKLE REDUCTION OF SENTINEL-1 GRD IMAGES}

\name{Inès Meraoumia$^\dagger$\thanks{This work has been partly funded by Futur and Rupture Program of Institut Mines-Télécom and by ANR (the French National Research Agency) and DGA (Direction Générale de
l’Armement) under ASTRAL project ANR-21-ASTR-0011.}, Emanuele Dalsasso$^\dagger$, Loïc Denis$^\ddagger$, Florence Tupin$^\dagger$}
\address{$^1$LTCI, Télécom Paris, Institut Polytechnique de Paris, Palaiseau, France \\ $^\ddagger$Univ Lyon, UJM-Saint-Etienne, CNRS,
Institut d Optique Graduate School,\\ Laboratoire Hubert Curien UMR
5516, F-42023, SAINT-ETIENNE, France\\ }

\let\OLDthebibliography\thebibliography
\renewcommand\thebibliography[1]{
  \OLDthebibliography{#1}
  \setlength{\parskip}{2pt}
  \setlength{\itemsep}{0pt plus 0.3ex}
}

\begin{document}
\maketitle
\begin{abstract}
Reducing speckle and limiting the variations of the physical parameters in Synthetic Aperture Radar (SAR) images is often a key-step to fully exploit the potential of such data. Nowadays, deep learning approaches produce state of the art results in single-image SAR restoration. Nevertheless, huge multi-temporal stacks are now often available and could be efficiently exploited to further improve image quality. This paper explores two fast strategies employing a single-image despeckling algorithm, namely SAR2SAR \cite{dalsasso2020sar2sar}, in a multi-temporal framework. The first one is based on Quegan filter \cite{Queg-01} and replaces the local reflectivity pre-estimation by SAR2SAR. The second one uses SAR2SAR to suppress speckle from a ratio image encoding the multi-temporal information under the form of a "super-image", i.e. the temporal arithmetic mean of a time series. Experimental results on Sentinel-1 GRD data show that these two multi-temporal strategies provide improved filtering results while adding a limited computational cost. 

\end{abstract}
\begin{keywords}
SAR imaging, speckle reduction, deep learning, multi-temporal series
\end{keywords}
\section{Introduction}
\label{sec:intro}

The success of SAR data for Earth Observation (EO) is confirmed by the improvements brought by the launch of new sensors in the recent years, with unprecedented spatial and temporal resolutions. Despite that, SAR data exploitation remains difficult. Indeed, even in physically homogeneous areas, the measured signal presents strong fluctuations inherent to coherent imaging systems and referred to as  speckle. 

Although being well modeled by Goodman model 
as a multiplicative noise \cite{Good-76}, speckle reduction remains challenging. In the past few years, unprecedented advances have been made thanks to deep learning based approaches \cite{chierchia2017sar,dalsasso:ujm-03270455}. Due to the lack of available databases for such a task, different strategies have been developed to train Convolutional Neural Networks (CNN) \cite{denis:ujm-03123042}. Speckle-free reference images can be obtained by multi-temporal averaging of SAR stacks or by relying on optical images. The first choice is more adapted since the specificities of SAR signals are respected (strong backscattering targets, bright lines,...). The simulation of a speckle-corrupted image is then usually done by multiplying the speckle-free reference with synthetic speckle, relying on Goodman's model. Although widely used, this strategy does not take into account the SAR system processing steps (over-sampling, spectrum apodization) which lead to spatially-correlated speckle \cite{dalsasso2020handle}. Thus, the community moved towards self-supervised learning algorithm (spurred by \cite{lehtinen2018noise2noise}) that directly exploit real SAR images. SAR2SAR \cite{dalsasso2020sar2sar} uses image pairs to obtain independent speckle-corrupted images, provided an adequate compensation of changes is performed. A fully unsupervised strategy exploiting a single image can be built either using a blind-spot technique like in \cite{molini2020speckle2void}, or using the complex nature of the SAR signal with MERLIN framework \cite{dalsasso:ujm-03270455}.

Nowadays, more and more multi-temporal data are available. This is particularly true 
since the launch of the Sentinel-1 satellite mission by ESA (European Space Agency) in the Copernicus program and the policy of freely available images. 

Incorporating multi-temporal information to help reducing the fluctuations of SAR data is an old subject, started with ERS-1 images \cite{Queg-01}. While providing a drastic improvement when the scene has not changed between the multi-temporal acquisitions, when changes occur they have to be taken into account carefully to avoid the creation of spurious structures. 
Many multi-temporal filters have been proposed in the past years, like \cite{Le-14,su:hal-01185740,cherc-17,lobry:ujm-01376896} to cite a few. 

This paper studies two fast strategies to combine the best of single-image CNN despeckling while exploiting multi-temporal redundancy. It especially focuses on Sentinel-1 GRD data, which are widely used for operational programs (such as forest monitoring). Please note that a temporal stack is constituted by finely registered images acquired on the same orbit with the same configuration (either ascending or descending). As a despeckling method, we consider the adaptation of SAR2SAR for GRD images proposed in \cite{gasnier:hal-03129006}, and presented in section 1. The two multi-temporal frameworks considered, Quegan \cite{Queg-01} and RABASAR \cite{rabasar}, are described in section 2. Then section 4 shows the benefits of employing SAR2SAR in a multi-temporal method through experimental results. 

\section{The SAR2SAR approach for GRD images}
\label{sec:sar2sar}
In this paper, we consider the multiplicative speckle model introduced by Goodman \cite{Good-76}. Denoting by $\intensity$ the observed intensity of the image, $\reflectivity$ the reflectivity that we aim to retrieve, and by $\speckle$ the speckle component, we have: $\intensity = \reflectivity \times \speckle$.

Within SAR2SAR, a CNN is trained to estimate $\reflectivity$ at each pixel using pairs of co-registered SAR images acquired at different dates. To take into account possible changes, a compensation is applied: the model is pre-trained on reference images corrupted with synthetic speckle and used to pre-estimate the reflectivities. Then, the network is fine-tuned on multi-temporal SAR stacks, using the pre-trained model to account for changes. Being trained on real SAR images, the network gains robustness to speckle spatial correlation. For a full description of the training strategy, the reader can refer to \cite{dalsasso2020sar2sar}.

This framework has first been developed on single-look Sentinel-1 images. An adaptation to GRD data, presenting a different number of looks and a spatially varying speckle correlation, is presented in \cite{gasnier:hal-03129006}. 
This network, available at {\url{https://gitlab.telecom-paris.fr/RING/SAR2SAR}},  will be used in the following multi-temporal strategies.

\section{Two fast strategies for multi-temporal speckle reduction}
\label{sec:mtsar2sar}
In this section we propose two efficient adaptations that turn single-image despeckling methods into multi-temporal filters. 

\subsection{Adaptation of Quegan filter}

Quegan filter \cite{Queg-01} is a powerful yet simple approach, when taking the approximate version, to denoise multi-temporal stacks. Indeed, if the temporal correlation between the images is neglected, it boils down to an average of change compensated images. Let us denote by $\intensity_t$ and $\reflectivity_t$ the intensity and reflectivity of a specific date $t$, the filtering formula giving the reflectivity $\hat{\reflectivity _t}^{Q}$ of data $t$ is the following (with $T$ the number of dates of the stack):
\begin{equation}
 \hat{\reflectivity _t}^{Q}=\frac{1}{T}\sum_k {\hat{\reflectivity_t}} \frac{\intensity_k}{\hat{\reflectivity_k}} 
\end{equation}
The change compensation between date $t$ and a date $k$ of the multi-temporal stack is done thanks to an estimation of the reflectivity of each date $k$ denoted by $\hat{\reflectivity_k}$. This is a "chicken-and-the-egg" problem since in case of a perfect knowledge of $\hat{\reflectivity_k}$, the multi-temporal denoising would not be needed. Therefore, the better this estimation, the better the multi-temporal denoised result. In practice, in the original paper, it is proposed to evaluate these estimates by local averages of the intensity values around the processed pixel for each date (corresponding to a local spatial multi-looking). Of course this estimation leads to a loss of resolution, for instance blurring strong targets and edges, which impacts the global multi-temporal result. 

A simple improvement is thus to replace these estimates by more efficient estimations for instance provided by SAR2SAR-GRD. Experimental results are shown and discussed in Section 4 and Fig.\ref{fig:quegan}.

\subsection{Adaptation of RABASAR filter}
A second strategy is the ratio based approach proposed in \cite{rabasar}. The idea is to create a "super-image" $s$ (obtained by a temporal arithmetic mean) and to denoise a residual image corresponding to the ratio between a specific data $t$ and this "super-image" $\ratio_t=\frac{\intensity_t}{s}$. In this case, there is no change compensation before averaging as in Quegan strategy, but it is claimed that the residual image is easier to denoise than the original one as it has an improved stationarity. Finally, the denoised estimation of $\reflectivity_t$ is retrieved by multiplying $\hat{\ratio}_t$ with the super-image $\superimage$:
\begin{equation}
   \hat{\reflectivity}^{R}_t = \hat{\ratio}_t \times \superimage  
\end{equation}

As already presented in \cite{dalsasso:hal-03129020}, we propose to use SAR2SAR for the denoising of the ratio image $\ratio_t$ of date $t$. In this paper we are moslty interested in GRD data and therefore we have to take into account the multi-look processing applied on these data. As mentioned in section 2, this is done by using the network presented in \cite{gasnier:hal-03129006} which was trained on GRD images. 

To further improve the results, it is possible to use a denoised "super-image" to limit the residual fluctuations. Indeed, depending of the available number of dates in the stack, the "super-image" can still present some fluctuations, even in temporally stable areas.

\begin{figure*}[p]
\centerline{\includegraphics[width=\textwidth]{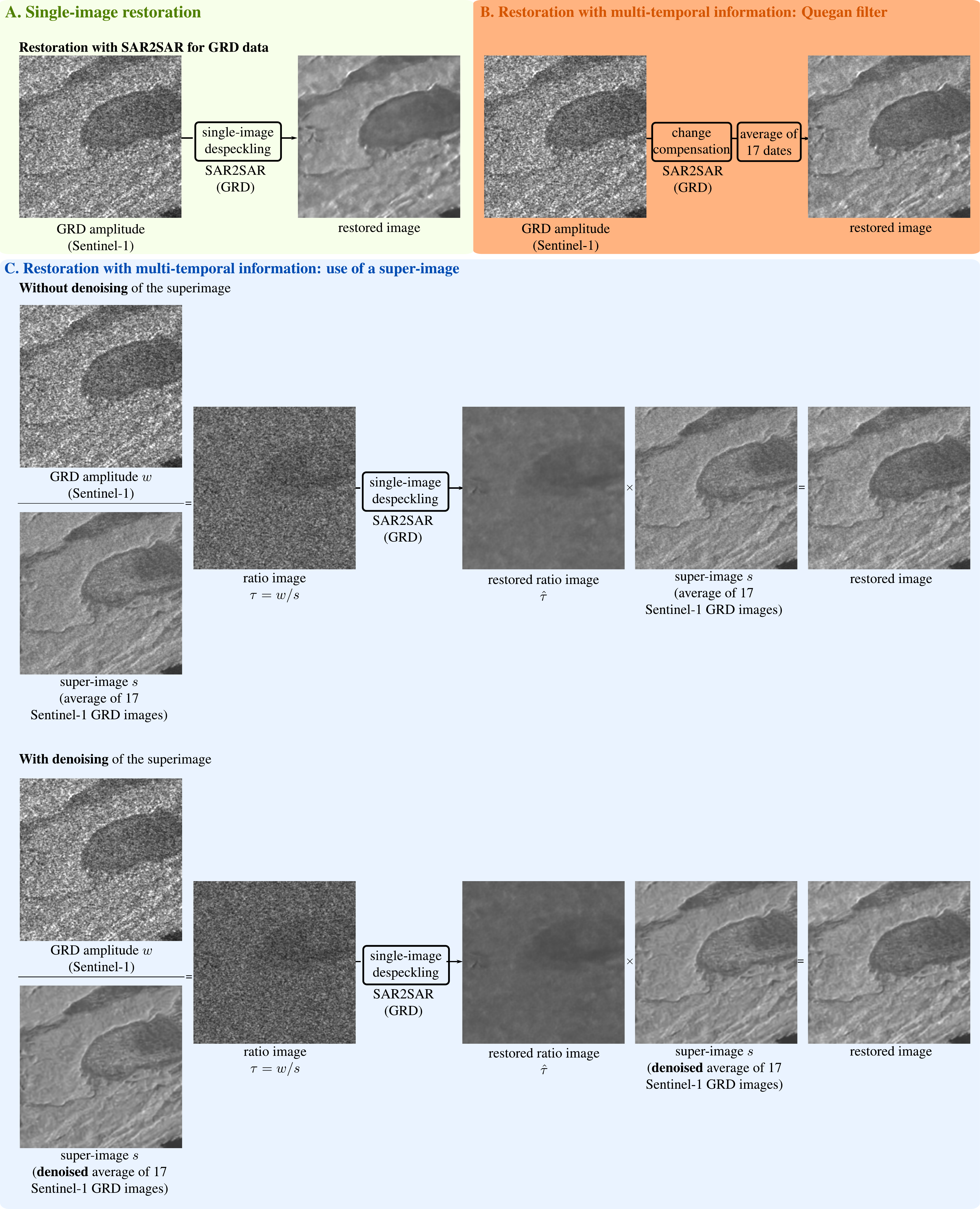}}\vspace*{-\baselineskip}
\caption{Comparison of several despeckling strategies for single-image and multi-temporal processing.}
\label{fig:allres}
\end{figure*}

\section{Experimental results}
\label{sec:exp}

We illustrate the results of the different approaches on figure \ref{fig:allres}. The stack is constituted by 17 Sentinel-1 GRD images acquired around the town of Mallacoota, Australia. 

The first row illustrates a single image deep-learning based filtering (on the left, block A.) and its introduction in the  multi-temporal framework of Quegan (block B.). It can be observed that the result is improved with a much better preservation of edges, lines and points, as expected. 
The second and third rows illustrate the RABASAR strategies with a deep-learning based filtering of the ratio image. 
The results obtained are very similar to the one given by Quegan approach. Besides, the use of a denoised "super-image" has a reduced impact for long time series with limited changes, and the two results (second and third rows) are quite close. 

\begin{figure}[htpb]
    \centering
    \includegraphics[width=0.4\textwidth]{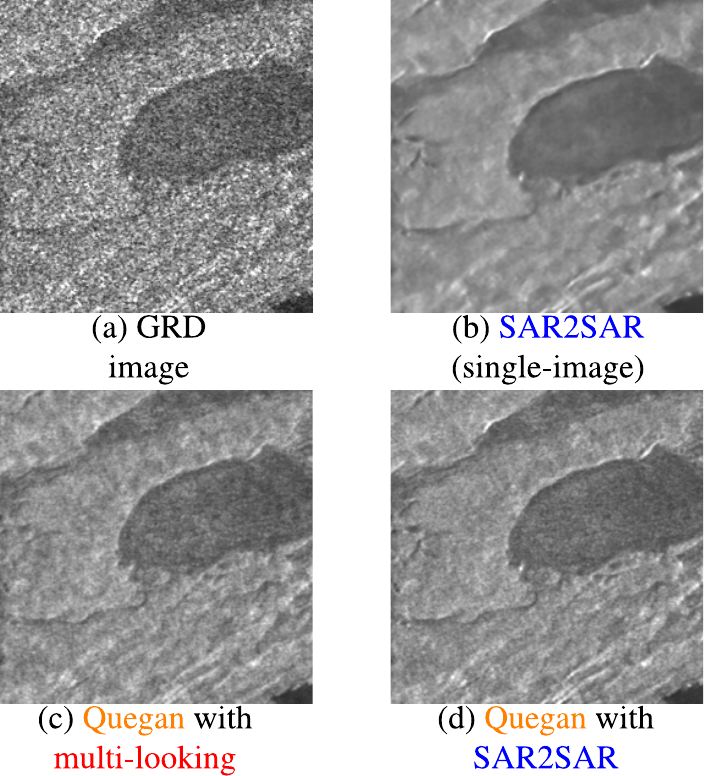}
    \caption{Details preservation and image quality is improved when SAR2SAR is integrated within Quegan filter (fig.(d)). On the one hand, single-image restoration results in a loss of fine structures (fig.(b)). On the other hand, the Quegan filter requires a high-quality speckle reduction algorithm: indeed, pre-estimating the reflectivites with a spatial averaging blurs some of the image details (fig.(c)).}
    \label{fig:quegan}
\end{figure}

Figure \ref{fig:quegan} shows more in details the improvement brought by integrating SAR2SAR within the multi-temporal Quegan filter. Replacing the spatial multilooking by SAR2SAR results in a better preservation of details.

\section{Conclusion}
\label{sec:concl}
The introduction of deep learning despeckling methods in multi-temporal frameworks allows a strong improvement even using simple multi-temporal strategies such as change-compensated averaging or ratio denoising. 

A major interest of these approaches is that they can be implemented with a low computational complexity by pre-computing the averages of multi-temporal data and storing them for the processing of many newly acquired dates. Besides the updating of these averages can be done very easily. 

We are convinced that these denoising frameworks can be very useful for operational applications exploiting GRD Sentinel-1 images at low computational cost. A demo code will be made available upon publication. It will allow to reproduce this paper's results, as well as applying the described approaches on more GRD SAR stacks.

\section{Acknowledgements}
The authors would like to thank Lucie Jandet (Sorbonnes University and Mines ParisTech), for producing some of the experimental results.

\small
\bibliographystyle{IEEEbib}
\bibliography{paper_2022}

\end{document}